\documentclass[conference]{IEEEtran}
\IEEEoverridecommandlockouts
\usepackage{amsmath,amssymb,amsfonts}
\usepackage{algorithmic}
\usepackage{graphicx}
\usepackage{textcomp}
\usepackage{xcolor}
\def\BibTeX{{\rm B\kern-.05em{\sc i\kern-.025em b}\kern-.08em
    T\kern-.1667em\lower.7ex\hbox{E}\kern-.125emX}}

\usepackage[style=ieee]{biblatex}
\usepackage{bm}
\usepackage{enumitem}
\usepackage{multirow}
\usepackage{booktabs}

\begin{document}

% \title{Conference Paper Title*\\
\title{Foundational Models for Fault Diagnosis of Electrical Motors}

\author{\IEEEauthorblockN{Sriram Anbalagan}
\IEEEauthorblockA{\textit{Department of Applied Mechanics} \\
\textit{IIT Madras}\\
Chennai, Tamil Nadu, India \\
am21s013@smail.iitm.ac.in}
\and
\IEEEauthorblockN{Deepesh Agarwal}
\IEEEauthorblockA{\textit{Department of Electrical and Computer Engineering} \\
\textit{Kansas State University}\\
Manhattan, Kansas, USA \\
deepesh@ksu.edu}
\and
\IEEEauthorblockN{Balasubramaniam Natarajan}
\IEEEauthorblockA{\textit{Department of Electrical and Computer Engineering} \\
\textit{Kansas State University}\\
Manhattan, Kansas, USA \\
bala@ksu.edu}
\and
\IEEEauthorblockN{Babji Srinivasan}
\IEEEauthorblockA{\textit{Department of Applied Mechanics} \\
\textit{IIT Madras}\\
Chennai, Tamil Nadu, India \\
babji.srinivasan@iitm.ac.in}
}

\maketitle

% \begin{abstract}
% This document is a model and instructions for \LaTeX.
% This and the IEEEtran.cls file define the components of your paper [title, text, heads, etc.]. *CRITICAL: Do Not Use Symbols, Special Characters, Footnotes, 
% or Math in Paper Title or Abstract.
% \end{abstract}

\begin{abstract}
A majority of recent advancements related to the fault diagnosis of electrical motors are based on the assumption that training and testing data are drawn from the same distribution. However, the data distribution can vary across different operating conditions during real-world operating scenarios of electrical motors. Consequently, this assumption limits the practical implementation of existing studies for fault diagnosis, as they rely on fully labelled training data spanning all operating conditions and assume a consistent distribution. This is because obtaining a large number of labelled samples for several machines across different fault cases and operating scenarios may be unfeasible. In order to overcome the aforementioned limitations, this work proposes a framework to develop a foundational model for fault diagnosis of electrical motors. It involves building a neural network-based backbone to learn high-level features using self-supervised learning, and then fine-tuning the backbone to achieve specific objectives. The primary advantage of such an approach is that the backbone can be fine-tuned to achieve a wide variety of target tasks using very less amount of training data as compared to traditional supervised learning methodologies. The empirical evaluation demonstrates the effectiveness of the proposed approach by obtaining more than 90\% classification accuracy by fine-tuning the backbone not only across different types of fault scenarios or operating conditions, but also across different machines. This illustrates the promising potential of the proposed approach for cross-machine fault diagnosis tasks in real-world applications. \footnote{This work has been submitted to the IEEE for possible publication. Copyright may be transferred without notice, after which this version may no longer be accessible.}
\end{abstract}


\section{Introduction}

Electrical motors are considered to be the workhorses of industries. They deliver reliable and efficient operation across diverse range of applications including commercial, industrial, aerospace, computer systems, robotics and defense \cite{liu2020critical}. However, these motors are susceptible to a variety of faults that can disrupt their performance, compromise system safety, and lead to unexpected downtime \cite{9299395}. Efficient and timely detection of faults in electrical motors is crucial for ensuring their optimal operation, minimizing maintenance costs, and preventing catastrophic failures \cite{9487715}. Robust fault diagnosis and degradation analysis methods for electrical motors have been developed over years of study and engineering work. The goal of fault diagnostics is to recognize and categorize various fault types, from electrical faults like inter-turn fault, stator winding failures to mechanical defects like bearing wear, shaft misalignment and broken rotor bars. The availability of machine condition monitoring data enhanced the development of many fault diagnosis methods. However, most of the existing approaches assume that the distribution of data is the same across training and testing datasets \cite{zhu2022review}, and the model performs poorly when encountered with different fault types and varying severity levels. Based on the application domains, the electrical motors are designed to operate in wide variety of dynamic environments. Consequently, the distributions of the data in subsequent testing scenarios are expected to change from those of the data used in training the model \cite{li2022perspective}. Even though a large amount of condition monitoring data is collected, the manual annotation of data is costly, error-prone and labour-intensive \cite{zhu2022review}. In order to overcome the aforementioned drawbacks related to differences in distributions of training and testing subsets and limited availability of the labeled samples, this work proposes a framework to develop foundational model for fault diagnosis of electrical motors. 

The proposed framework comprises of two key steps: (i) building the backbone model; (ii) fine-tuning the backbone to achieve specific objectives. A neural network-based backbone is trained to extract high-level features via self-supervised learning approach, after which it is fine-tuned to capture finer details using only a limited amount of labelled data. The key advantage of this approach is that fine-tuning the backbone enables the model to adapt and perform effectively across a diverse range of target tasks, requiring significantly less training data than conventional supervised learning methodologies. This efficiency makes the proposed approach particularly valuable for real-world applications where labelled data may be scarce or expensive to acquire. The proposed foundational model addresses key challenges in building a real-time fault diagnosis model, such as transferability and adaptability, limited fault coverage, sensitivity to sensor noise, and computational capacity. By incorporating fine-tuning, the model overcomes the limitations of traditional transfer learning, making it versatile and effective in handling diverse fault diagnosis tasks across various industrial applications. Moreover, the ability of the model to work with minimal labeled samples enhances its practicality and usability in real-world scenarios, where data acquisition and labeling can be resource-intensive.
% The backbone model will effectively learn the meaningful representation from the labelled data, which is used to train the backbone. The novel fine-tuning approach will enable the application of the backbone model to multiple target tasks with limited labelled samples.  

% The backbone model is a 15-layer 1D CNN network employed for self-supervised learning, effectively learning meaningful representations from labeled data. In the novel proposed approach, the developed backbone model undergoes fine-tuning, enabling its applicability to multiple target tasks with limited labelled samples.

%This work proposes the Foundational model, a novel approach for robust fault detection using raw condition monitoring data.

\subsection{Related Work}
%MFD solved decades back, say the existing techniques that were used to solve this problem. cite deepesh paper also. 
%initially sp, ml, dl, TL, SSL, 
In the context of modern industry, machines and equipment are continuously evolving, aiming for higher precision, efficiency, automation, and complexity. However, these advancements also come with an increased risk of breakdowns and accidents. As a result, electrical motor fault diagnosis becomes paramount in ensuring the safety and reliability of industrial equipment. Over the last decade, significant efforts have been dedicated to developing efficient algorithms and innovative approaches to achieve superior diagnostic performance. The existing approaches used for electrical motor fault diagnosis can be broadly grouped into three categories. Firstly, advanced signal processing techniques \cite{chen2016wavelet}, \cite{wang2016spectral} are used to identify the types and locations of faults in machines. However, these methods heavily rely on specialized knowledge that is often lacking among maintenance personnel in engineering scenarios. Moreover, the diagnostic outcomes produced by signal processing techniques can be highly specialized and challenging for machine users to comprehend. As a result, contemporary industrial applications seek fault diagnosis methods that can automatically recognize the health status of machines.

Secondly, with the help of probabilistic machine-learning and neural network-based techniques such as ANN, support vector machine (SVM), random forest (RF), \cite{agarwal2020fault}, convolutional neural network (CNN), stacked autoencoder (SAE), deep belief network (DBN), and deep neural network (DNN) are the most popular approaches that have been widely deployed in electrical motor fault diagnosis \cite{qiu2019deep} \cite{9432177}. However, these methods assume that the training and testing datasets are from the same distribution, making these diagnostic models less robust in varying working conditions of motors.

Thirdly, transfer learning, which is a branch of machine learning, emphasizes acquiring common knowledge from one or more related but different application scenarios. It aids AI algorithms in achieving enhanced performance for a specific application of interest, making it a promising electrical motor fault diagnosis methodology \cite{li2022perspective}, \cite{10042467}. However, the effectiveness of transfer learning in generalizing to the target domain and maintaining diagnostic accuracy may reduce when there is a substantial difference between the source and target domains. The quality and size of the source domain dataset play a crucial role in transfer learning, and inadequate source data can lead to decreased performance. 

Existing fault diagnosis models have trouble adjusting to new motor configurations or operating circumstances that weren't covered by the training dataset \cite{li2022perspective}. Model transferability to various motor kinds, sizes, or manufacturers is still difficult and necessitates complex retraining or fine-tuning work.  %final lines of related work.
This article presents a novel approach to develop a foundational model-based electrical motor fault diagnosis, aiming to diagnose various faults in electrical motors under different working conditions and across different motor types.

\subsection{Contributions}

This work, for the first time, presents a framework for developing foundational models for fault diagnosis of electrical motors. This problem is approached as a two-step process: 
1) Developing a backbone model to learn high-level features via self-supervised learning.
2) Fine-tuning approach to capture the finer details using a limited amount of labelled data. 
The foundational model is evaluated based on the following aspects: (a) \textit{Expressivity} - the capacity to acquire and assimilate real-world data efficiently; (b) \textit{Scalability} - to manage large volumes of high-dimensional data
effectively; and (c) \textit{Generalizability} - the ability to work in varying environmental conditions. The experimental evaluation reveals that the proposed model demonstrates remarkable performance in diagnosing electrical motor faults by achieving the above attributes of the foundational model. Such a study based on foundational models has not been presented yet in the literature related to fault diagnosis of electrical motors and is a novel contribution of this work.

The remainder of this article is structured as follows. Section \ref{sec:Preliminary} briefly introduces electrical motor faults and the concept of foundational models. The proposed framework for developing foundational model for fault diagnosis of electrical motors is presented in Section \ref{sec:Methodology}. The empirical evaluation is conducted and the results are discussed in Section \ref{sec:Results}. The article ends with concluding remarks in Section \ref{sec:conclusions}.

\section{Preliminaries} \label{sec:Preliminary}

\subsection{Faults in Electrical Motors}

An electrical motor has mechanical components like stator, rotor, bearings, and electrical components such as windings and end rings. These components work together to facilitate the motor's operation. Electrical motors are engineered to operate under varying industrial and environmental conditions, subjecting them to a wide range of stresses \cite{9299395}. These stresses contribute to the emergence of various faults within the electrical motor, broadly categorized as mechanical and electrical faults. Among the major faults encountered in electrical motors are bearing faults, shaft misalignment, rotor unbalance, and inter-turn short circuit faults. According to the existing literature, the majority of electrical motor failures are attributed to mechanical faults \cite{choudhary2019condition}. In this paper, we address a majority of the mechanical faults that occur in electrical motors, including bearing faults of different sizes considered at various locations (i.e., inner-race, ball, outer-race), rotor unbalance faults and shaft misalignment faults at multiple severity levels.

%Electrical motors are designed to function in demanding industrial and environmental conditions, experiencing a wide range of stresses. These stresses lead to various faults in the electrical motor, which can be classified as mechanical and electrical faults. Bearing, Shaft misalignment, Rotor Unbalance, and Inter-turn Short circuit faults are the major faults that occur in electrical motors. According to the literature, failure of electrical motors is majorly due to mechanical fault \cite{choudhary2019condition}. In this paper, we consider almost all types of mechanical faults of electrical motors to train the foundational model. %Later, it can be used for several target tasks.     

\subsection{Foundational Models}

Foundational models are technically enabled by transfer learning and scale \cite{bommasani2021opportunities}. The principle behind transfer learning is applying the knowledge gained from one work to another. Pretraining is the prevalent method of transfer learning in deep learning. A model is trained on a surrogate task and tuned to fit the downstream task of interest. Scale makes the foundational model powerful. Foundational models have been recently implemented in both natural language processing (NLP) and computer vision tasks. For instance, the masked language modelling task in Bidirectional Encoder Representations from Transformers (BERT) \cite{kenton2019bert} involves predicting a missing word within a sentence based on its context. However, the true impact of these foundational models on NLP lies not solely in their raw generation capabilities but rather in their remarkable versatility and adaptability. A single foundational model can be effectively customized in multiple ways to perform various linguistic tasks, making them invaluable tools in the field of NLP.  
However, to the best of authors' knowledge, there is no existing implementation of a foundational model in the field of fault diagnosis of electrical motors. The following section presents a detailed description of our proposed foundational model scheme for fault diagnosis of electrical motors.

\section{Methodology} \label{sec:Methodology}

\begin{figure}
    \centering
    \includegraphics[width=\linewidth]{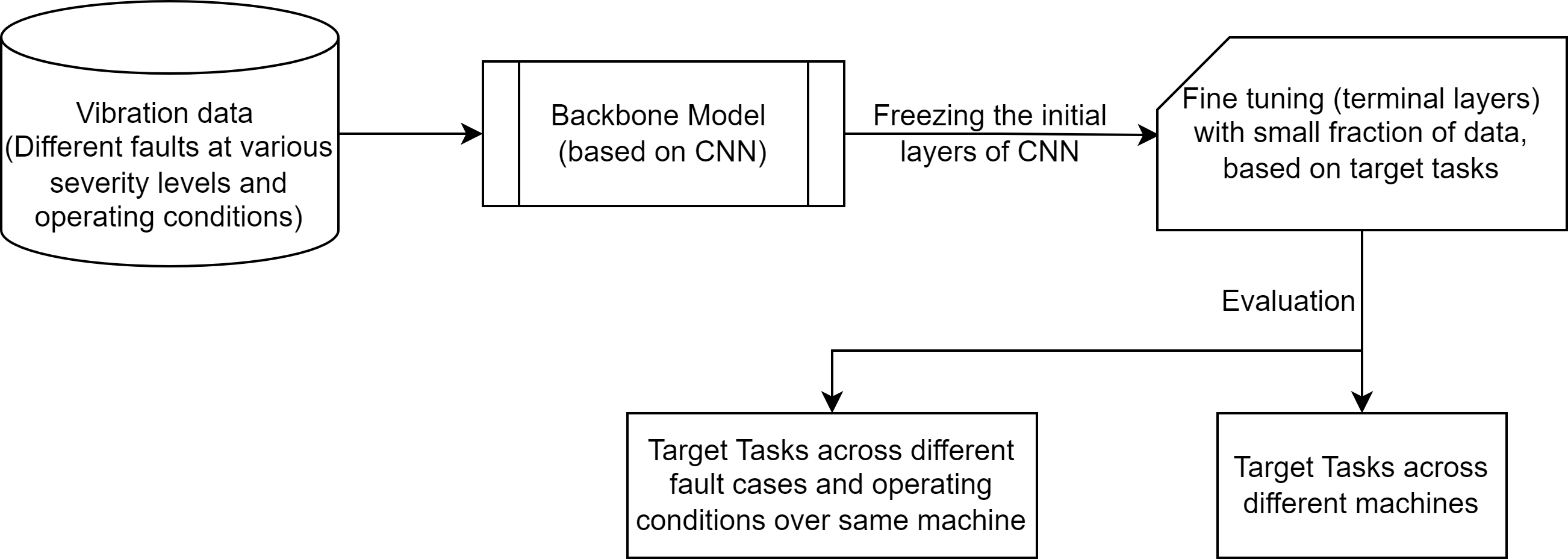}
    \caption{Proposed framework for developing foundational models for fault diagnosis of electrical motors.}
    \label{fig:prop_frmw}
\end{figure}

The proposed framework for developing a foundational model for fault diagnosis of electrical motors is illustrated in Figure \ref{fig:prop_frmw}. It is approached as a two-step process: 
\begin{enumerate}
    \item A CNN-based backbone model is developed to learn higher-level representations encompassing all potential electrical motor faults.
    \item Based on the requirements of individual target tasks, the backbone is fine-tuned to achieve specific objectives.
\end{enumerate}

This systematic approach equips the foundational model with the ability to effectively diagnose various electrical motor faults with superior performance in a wide range of scenarios using small fractions of data. All the elements involved in building the foundational models are discussed next. 
%Training the backbone model, defining the target tasks and fine-tuning the backbone model is the process involved in the foundational model discussed in this section.
%The foundational model learns the higher-level representation of all types of possible electrical motor faults using the 1D CNN architecture. The novel fine-tuning method leverages the model performance in several target tasks within the same motor and even target tasks in another motor. 
%The foundational model is pre-trained using a labelled dataset consisting of all possible electrical motor faults, which is then fine-tuned for multiple target tasks. During fine-tuning, we use 5\% to 25\% of the labelled dataset for that target task and can achieve higher accuracy with less time and effort. 
		% 	\begin{figure}[htbp]
		% 	\centerline{\includegraphics[width=0.9\linewidth, height=0.3\textwidth ]{./images/FM2_flow.png}}
			
		% 	\caption{General pipeline of the Electrical Motor fault diagnosis based on the proposed method (TT = Target task in the figure).}
		% 	\label{fig:flowdiag}
		% \end{figure}

\subsection{Training the Backbone Model}

The backbone model serves as the fundamental architecture, which is trained using data corresponding to various mechanical faults as well as healthy scenarios over constant speed and varying speed conditions. A recently published dataset \cite{jung2023vibration} is used for training the backbone model. It consists of vibration, current, temperature and acoustic data for different faults, including bearing faults (inner and outer races) at constant and variable speeds, shaft misalignment faults, and rotor unbalance faults at a constant speed. A 1D CNN architecture is employed to develop the backbone model, known for its superior performance with raw time series data compared to other networks \cite{wang2021bearing}. The proposed architecture comprises of 15 convolutional layers, a global average pooling layer, and a dense layer. Each convolutional layer consists of 64 filters with a kernel size of 3, employing LeakyReLU as the activation function. The input samples are processed through the initial CNN layer, and this architecture is maintained throughout the subsequent 14 layers. The final layer is a global average pooling layer, followed by a dense layer serving as the output layer with the softmax activation function. We utilize the Adam optimizer for optimisation, while the loss function is chosen as sparse categorical cross-entropy. These choices contribute to the overall effectiveness and efficiency of our model.

\subsection{Defining Target Tasks} \label{subsec:define TT}

The target tasks considered in this work are divided into two groups, to evaluate the performance of foundational model across (a) fault cases and operating conditions within same machine; and (b) different machines. A hierarchical approach is followed to design the tasks in first group. The initial few tasks are aimed at categorizing the samples into healthy vs. faulty. Once the fault is detected, the subsequent target tasks are designed to diagnose the type, location and severity of faults. Such a hierarchical approach enables a comprehensive fault diagnosis, offering detailed insights into the nature and location of potential faults. This is applied to both constant speed and varying speed conditions. The list of target tasks are tabulated in Table \ref{tab:target task}.
The second group of target tasks are designed to evaluate the performance of foundational model across different machines. Specifically, the data from a different motor (i.e., other than the one used for training the backbone model) is used for fine-tuning. Here, the model is tasked to perform bearing fault diagnosis at different speeds and loading conditions. These diverse set of target tasks validates the scalability and adaptability across varying motor datasets, showcasing its versatility to diagnose faults in different scenarios. Furthermore, we also subject the vibration measurements to white Gaussian noise in order to mimic measurement uncertainties and execute the same set of target tasks enlisted in Table \ref{tab:target task}. This further helps to assess the generalizability and robustness of the proposed approach.

\begin{table}[]
		\centering 
		\caption{ Target Task performed with foundational model}
		\label{tab:target task}
\resizebox{\columnwidth}{!}{%
\begin{tabular}{rl}
\toprule
\multicolumn{1}{c}{S.No.} & \multicolumn{1}{c}{Target Task}
\\ 
\midrule
1                         & \begin{tabular}[c]{@{}l@{}}Fault Detection at constant speed: Classification-Healthy vs. Faulty\end{tabular}                                                         \\
\midrule
2                         & \begin{tabular}[c]{@{}l@{}}Fault Detection at variable speed: Classification-Healthy vs. Faulty\end{tabular}                                                         \\
\midrule
3                         & \begin{tabular}[c]{@{}l@{}}Fault Diagnosis at constant speed: Classification- Bearing Fault\\ vs Shaft Misalignment vs Rotor Imbalance\end{tabular}                 \\
\midrule
4                         & \begin{tabular}[c]{@{}l@{}}Bearing Fault Diagnosis at constant speed:Classification - \\Inner Race vs. Outer Race\end{tabular}                                         \\
\midrule
5                         & \begin{tabular}[c]{@{}l@{}}Fault Diagnosis at variable fault size: Classification-Healthy vs. \\Bearing Fault vs Shaft Misalignment vs Rotor Imbalance\end{tabular} \\
\midrule
6                         & \begin{tabular}[c]{@{}l@{}}Bearing Fault Diagnosis at variable fault size: Classification- \\Inner Race vs Outer Race\end{tabular}\\
\bottomrule
\end{tabular}
}
\end{table}

\subsection{Fine-tuning the Backbone Model} \label{subsec:fine-tune}

In transfer learning, fine-tuning refers to the process of adapting a pre-trained neural network model to a new task or domain. The information and characteristics learnt from a pre-trained model are borrowed instead of training a model from scratch, saving a significant amount of time and eliminating the need of massive training data. There are various phases involved in fine-tuning. First, a pre-trained model is chosen as the starting point. Pre-trained models are often trained on a big dataset and relevant tasks. Next, the weights of the pre-trained model are frozen to avoid any changes during the preliminary training stage. The model maintains the learnt features and prevents overfitting on the sparse data of the new job by freezing the weights. The pre-trained model then gets augmented by a fresh set of fully connected layers or a few top layers. The pre-trained characteristics of the model must be adjusted to the new job or domain via these newly added top layers. The new layers are fine-tuned using a smaller dataset tailored to the new task during training. By fine-tuning, the model adapts to the unique properties of the new task or domain while inheriting and transferring the information and representations learnt during the pre-training phase. It is expected to achieve stellar performance on the target job while greatly reducing the training time and data requirements.

In this work, we execute fine-tuning by focusing on initial and final dense layers of the pre-trained model. In the proposed model, we unfreeze the first three layers out of the fifteen layers utilized in the 1D CNN architecture, enabling them to be trainable during fine-tuning for various target tasks. By fine-tuning these frontal layers, we observed improved performance across various target tasks that we specifically designed for the machine dataset on which the foundational model was originally pre-trained. This unique fine-tuning approach, distinct from traditional transfer learning techniques, expands the capabilities of the foundational model from intra-machine fault diagnosis to inter-machine fault diagnosis with exceptional accuracy, even when utilizing a minimal amount of labelled samples. Section \ref{sec:Results} provides a comprehensive account of the robust performance of the foundational model  when applied to intra- and inter-machine fault scenarios.

\section{Results} \label{sec:Results}

In order to demonstrate the practical utility of the proposed foundational model in electrical motor fault diagnosis, we conduct a thorough evaluation by fine-tuning the backbone for multiple target tasks involving different fault cases over constant speed and variable speed conditions for same machine as well as other machines. We emphasize that the proposed framework achieves three key attributes, i.e., expressivity, scalability, and generalizability, which are essential for the development of a robust foundational model. In this work, we use three datasets from three distinct machines to train the backbone model and fine-tune it to perform various target tasks during intra- and inter-machine scenarios.
% To demonstrate the effective application of the foundational model for fault diagnosis in electrical motors, we evaluate the model in several target tasks within the same machine and different machine data. We highlight that our proposed foundational model meets the basic attributes such as expressivity, scalability, and generalizability required for the development of a foundational model.   
A recently published dataset \cite{jung2023vibration} is utilized to develop the backbone model. It consists of two parts: the first part includes vibration, acoustic, temperature, and driving current data collected under varying load conditions. The data comprises measurements under three loading conditions: 0 Nm, 2 Nm and 4 Nm. The sampling frequency for vibration, temperature, and driving current data was set at 25.6 kHz. This dataset contains 120 seconds of vibration data in normal states and 60 seconds in faulty states. The second part of the dataset focuses on vibration and current data acquired from the bearing, faults at different locations such as inner-race, outer-race, and ball. These data were collected under continuously varying speed conditions by modifying the motor speed between 680 and 2460 RPM.
In this work, we specifically consider the vibration data for bearing faults, shaft misalignment faults, rotor unbalance faults at a constant speed, and bearing faults at varying speeds.
%Each fault data at a constant speed is available for 60 seconds, while the variable speed data is available for 2100 seconds. We chose 60 seconds for each fault and normal condition data at constant and variable speeds to maintain an equal amount of data in each class.
The backbone model is trained by combining data from normal, inner race, outer race, shaft misalignment, rotor unbalance faults at constant speed, and normal, inner race, and outer race faults at variable speed conditions. 

We consider two additional machine datasets for performing target tasks on different machines: Jiangnan University (JNU) \cite{li2019school} and Case Western Reserve University Bearing Data Center (CWRU) \cite{case2018case}. The JNU dataset consists of three bearing vibration datasets captured at different rotating speeds, with data collected at a sampling frequency of 50 kHz. It contains one healthy state and three fault modes: inner race fault, outer race fault, and rolling element fault. The CWRU dataset consists of vibration signals collected from both normal bearings and damaged bearings with single-point defects. The data was collected under four different motor loads, and the sampling frequency varies between 12 kHz and 48 kHz. Within each working condition, single-point faults were intentionally introduced, with fault diameters of 0.007, 0.014, and 0.021 inches, targeting the rolling element, inner ring, and outer ring of the bearings, respectively. For this particular study, we focused on the data collected from the drive end, the sampling frequency used was set at 48 kHz, and the fault diameter of 0.014 inches was considered.

\subsection{Expressivity}

The expressivity of a foundational model is defined as its capacity to efficiently acquire and assimilate real-world data. It involves the ability of a model to recognize and understand the intricate relationships, characteristics, and patterns found in the data. A highly expressive model is better able to handle a variety of complex situations because it can more precisely reflect the nuances of real-world occurrences. An expressive model should be able to capture the minute fluctuations in vibration, acoustic, temperature, and driving current signals that indicate various fault situations in the motor system when used for fault diagnosis. To enable reliable detection and classification, it should be able to learn and comprehend the distinctive patterns linked to both normal functioning and various faults of electrical motors.

% \usepackage{multirow}
\begin{table}[]
		\centering 
		\caption{Expressivity: Fine-tuning the backbone for different target tasks within same machine}
		\label{tab: TTASK1}
\begin{tabular}{crrrrr}
\toprule
\multirow{2}{*}{Target Tasks} & \multicolumn{5}{r}{Percentage of data used for fine-tuning} \\
\cmidrule{2-6}
                              & 5\%        & 10\%       & 15\%      & 20\%      & 25\%      \\
                              \midrule
1                             & 87.53      & 88.65      & 88.85     & 92.71     & 95.33     \\
2                             & 67.95      & 72.47      & 73.40     & 84.58     & 85.04     \\
3                             & 86.11      & 91.62      & 97.80     & 98.20     & 99.46       \\
4                             & 91.54      & 93.06      & 97.37     & 98.51     & 99.33       \\
5                             & 88.70      & 89.44      & 90.31     & 92.16     & 93.74     \\
6                             & 81.58      & 94.17      & 96.48     & 98.24       & 99.39  \\
\bottomrule
\end{tabular}
\end{table}

To demonstrate the expressive capability of our proposed model, we fine-tuned the backbone model to target tasks outlined in Table \ref{tab:target task} using a machine dataset collected in real-time operation. The results of these target tasks, as presented in Table \ref{tab: TTASK1}, indicate that our model consistently achieves higher accuracy rates for most of the tasks, even when fine-tuned with much lesser labeled samples. A classification accuracy of more than 90\% is obtained for almost all the target tasks with just 15\% of the labeled samples. This illustrates the ability of the model to effectively capture and understand complex patterns in the data, leading to accurate classification and successful fault diagnosis.

\subsection{Scalability}

Scalability is a crucial aspect of the foundational model, enabling it to manage large volumes of high-dimensional data effectively. This capability is essential for accommodating growing datasets and successfully generalizing them to new and unseen instances. A scalable model possesses the capacity to efficiently process and analyze extensive datasets, facilitating real-time or near-real-time fault diagnosis in practical applications. Furthermore, it should demonstrate the ability to handle diverse machines, varying operating conditions, and different fault types while maintaining high performance and accuracy levels.

%TABLE FOR JNU dataset 
% \usepackage{multirow}
\begin{table}[]
		\centering 
		\caption{Scalability: Fine-tuning the backbone for different target tasks in different machine (JNU dataset)}
		\label{tab: JNU}
\begin{tabular}{crrrrr}
\toprule
\multirow{2}{*}{\begin{tabular}[c]{@{}c@{}} Motor Speed \end{tabular}} & \multicolumn{5}{r}{Percentage of data used for fine-tuning} \\
\cmidrule{2-6}
                                                                                                & 5\%        & 10\%      & 15\%      & 20\%      & 25\%       \\
                                                                                                \midrule
600 RPM                                                                                         & 94.15      & 95.32     & 96.36     & 99.24     & 99.68      \\
800 RPM                                                                                         & 92.66      & 94.40     & 95.45     & 99.16     & 99.50      \\
1000 RPM                                                                                        & 94.42      & 96.64     & 97.42     & 98.75     & 99.20 \\
\bottomrule
\end{tabular}
\end{table}

% \usepackage{multirow}
\begin{table}[]
		\centering 
		\caption{Scalability: Fine-tuning the backbone for different target tasks in different machine (CWRU dataset}
		\label{tab: cwru}
\begin{tabular}{crrrrr}
\toprule
\multirow{2}{*}{\begin{tabular}[c]{@{}c@{}} Loading Condition\end{tabular}} & \multicolumn{5}{r}{Percentage of data used for fine-tuning} \\
\cmidrule{2-6}
                                                                                                & 5\%        & 10\%       & 15\%       & 20\%      & 25\%     \\
                                                                                                \cmidrule{1-6}
0 HP                                                                                            & 95.66      & 97.24      & 98.00      & 99.33     & 99.58      \\
1 HP                                                                                            & 96.34      & 97.82      & 98.26      & 99.24     & 99.64      \\
2 HP                                                                                            & 95.50      & 96.89      & 98.12      & 99.02     & 99.34      \\
3 HP                                                                                            & 96.44      & 97.53      & 98.54      & 99.15     & 99.87 \\
\bottomrule
\end{tabular}
\end{table}

To demonstrate the scalability of our proposed model, we designed target tasks using datasets corresponding to different machines, namely JNU and CWRU. The fine-tuning procedure is executed based on the discussion presented in Section \ref{subsec:fine-tune}. It can be observed that a classification accuracy of more than 90\% is obtained for all the target tasks by using just 5\% of the labeled data for fine-tuning. This accomplishes the ability of the model to scale effectively across diverse machine datasets while maintaining its performance in terms of classification accuracy.

\subsection{Generalizability}

Generalizability is the capacity of a model to comprehend and extrapolate from complicated combinations of smaller components or attributes. It is essential for promoting successful generalization to new contexts and surroundings in the context of defect diagnostics. The ability of the model to reason about the combinations of various aspects or components and comprehend how they affect the overall state or condition being observed determines its level of generalizability. This enables the ability of the model to generalize its understanding and produce precise forecasts in unusual settings. The generalized model can manage fluctuations in the input and adapt to various environmental conditions.

% \usepackage{multirow}
\begin{table}[]
		\centering 
		\caption{Generalizability: Fine-tuning the backbone for different target tasks with noise}
		\label{tab: ttasknoisy}
\begin{tabular}{clllll}
\toprule
\multirow{2}{*}{Target Tasks} & \multicolumn{5}{r}{Percentage of data used for fine-tuning}                                                                               \\
\cmidrule{2-6}
                              & \multicolumn{1}{r}{5\%}   & \multicolumn{1}{r}{10\%}  & \multicolumn{1}{r}{15\%}  & \multicolumn{1}{r}{20\%}  & \multicolumn{1}{r}{25\%}  \\
                              \midrule
1                             & 83.42                     & 85.23                     & 86.15                     & 86.81                     & 87.60                     \\
2                             & 66.49                     & 67.04                     & 69.93                     & 77.71                     & 82.00                     \\
3                             & 80.73                     & 88.87                     & 91.63                     & 94.86                     & 97.56                     \\
4                             & 80.79                     & 91.48                     & 95.29                     & 98.40                     & 99.00                     \\
5                             & \multicolumn{1}{r}{81.12} & \multicolumn{1}{r}{89.44} & \multicolumn{1}{r}{90.31} & \multicolumn{1}{r}{92.16} & \multicolumn{1}{r}{93.74} \\
6                             & \multicolumn{1}{r}{81.58} & \multicolumn{1}{r}{94.17} & \multicolumn{1}{r}{97.48} & \multicolumn{1}{r}{98.42}   & \multicolumn{1}{r}{99.36}  \\
\bottomrule
\end{tabular}
\end{table}

To demonstrate the expressive capability of our proposed model, we fine-tuned the backbone model to target tasks enlisted in Table \ref{tab:target task} by introducing 10\%  white Gaussian noise to the raw vibration signal. The performance of the foundational model with the noisy signal is reported in Table \ref{tab: ttasknoisy}. It can be clearly observed that a classification accuracy of more than 90\% is obtained for almost all the target tasks with just 15\% of the labeled samples. This demonstrates the robustness of the model and highlights its ability to effectively handle and classify fault conditions even in the presence of noise.

\section{Conclusion} \label{sec:conclusions}

This work introduces a novel foundational model-based approach for fault diagnosis of electrical motors. It helps to overcome the limitations of different data distributions across working conditions and machines. In contrast to the existing fault diagnosis approaches, the proposed framework involves the development of a CNN-based backbone model that extracts higher-level features from the training dataset. The backbone is then fine-tuned to achieve specific objectives based on target tasks designed to perform diagnosis of different fault types and severity levels across multiple speeds and loading conditions. The proposed approach is evaluated based on three key attributes, namely, expressivity, scalability and generalizability, which are essential for the development of a robust foundational model. The empirical evaluation reveals that a good classification performance is obtained by fine-tuning the backbone with much lesser labeled samples as compared to supervised learning approaches. This approach gives excellent results, even when the fine-tuning step is executed for target tasks on another machine, thereby enhancing its scalability. 

The proposed model is evaluated in multiple target tasks by adding white Gaussian noise to the data. The results demonstrate the remarkable generalization ability of the model to handle different operating conditions effectively. Overall, the proposed foundational model-based approach offers a promising electrical motor fault diagnosis solution, accommodating variations in data distributions and achieving robust performance even with limited labelled samples. The scalability and generalization capabilities of the proposed model make it a valuable tool for real-world applications in fault diagnosis for electrical motors. The future extension of this work shall involve extensive testing across more target tasks and incorporating physics information within the foundational models.

\printbibliography

\end{document}